\documentclass{ws-procs975x65}

\begin{document}

\title{HIGH FREQUENCY GRAVITATIONAL WAVES GENERATION IN LASER PLASMA INTERACTION}

\author{Xavier RIBEYRE$^*$ and Vladimir TIKHONCHUK}

\address{Centre Lasers Intenses et Applications, UMR 5107, CNRS, Universit\'e Bordeaux 1, CEA,\\
Universit\'e Bordeaux 1, 351, cours de la Lib\'eration, 33405 Talence, France\\
$^*$Email: ribeyre@celia.u-bordeaux1.fr}

\begin{abstract}
Estimates of the emitted power and the metric perturbation of the gravitational waves generated in laser plasma interaction
are performed. The expected intensities are too low to be detected with the present day instruments.
\end{abstract}


\bodymatter
\section{Introduction}\label{aba:sec1}
Existence of gravitational waves is postulated in the theory of general relativity 
by Einstein \cite{Einstein16}, but no direct detection of 
such waves has been made so far. The best evidence of their existence is due to 
the work of 1993 Nobel laureates Taylor and Hulse \cite{Taylor79}. The search for gravitational 
waves (GW) radiated by extraterrestrial sources is carried out by large gravitational interferometer 
detectors LIGO and VIRGO \cite{Braginskii00}. These detectors address the low frequency spectral 
band between 10 Hz and 10 kHz. Recently, astrophysical sources of high frequency gravitational waves 
(HFGW: $\nu$ $>$ 100 kHz) were considered and this renew an interest for a GW Hertz experiment 
\cite{Rudenko03}, which consists in generation and receiving the GW signal on Earth. 
One of the first considerations of the gravitational Hertz experiment in laboratory was done by Weber 
\cite{Weber60} in a low frequency domain. In the high frequency domain the possibilities of GW generation 
in laboratory were considered by Rudenko \cite{Rudenko03} and by Chapline \cite{Chapline74}. In particular, 
Rudenko proposed a GW Hertz experiment associated with high power electromagnetic waves and acoustic impulsive 
or shock waves travelling and interacting with a non-linear opto-acoustic medium.

\section{GW generation with lasers}\label{aba:sec2}
The GW emission is proportional to that the third time derivative of the quadrupolar mass momentum 
$Q(t)$ is no zero. The power emitted by a local source of GW writes: $P_{GW}\simeq G \dddot Q^2/5c^5$ and 
the metric perturbation $h_{GW}$ is given by
 $h_{GW} \simeq G \ddot Q/R c^4$, where $G$, $c$ are the gravitational and light celerity constants and $R$ is the distance between
the GW generator and the detector.  

As lasers actually are the most powerful sources of electromagnetic energy on earth,
 we present here analytical estimates and numerical simulations of generation of 
the HFGW in interaction of high power laser pulse with a medium in different geometries. 
First, during the laser plasma interaction, a strong shock driven by the ablation pressure 
is generated in the bulk material. In this configuration, material is accelerated 
in the shock front and in the ablation zone. Because of a short laser pulse duration (ns) GW are generated 
in the GHz domain. During the laser interaction with a planar thick foil (more than $100 \mu$m thickness) 
the laser launch a shock with a velocity $V_s$, this shock accelerates the medium along the $z$ axis and produces a quadrupolar 
mass momentum $Q_{zz}$.
The time dependence of the quadrupolar mass momentum in shock is $Q_{zz}=S\rho_0 V_s^3 t^3$, where $S$ 
is the surface of laser focal spot.
The shock velocity $V_s\simeq \sqrt{P_s/\rho_0}$ depends on the ablation pressure $P_s$ and the material density $\rho_0$.
 Moreover, $P_s$ is directly connected to the laser intensity $I_L$ as: 
$P_s\simeq 112 (I_L) ^{2/3}$ (for the laser wavelength 0.35 $\mu$m), where $I_L$ is in PW/cm$^2$ and $P_s$ in Mbar \cite{AtzeniBook}.
With these  relations the radiated GW power reads: $P_{GW} [{\rm erg/s}]\simeq 7\times10^{-18} P_L^2/\rho_0$ and $h_{GW}\simeq3\times10^{-37}E_L/R\sqrt{\rho_0}$,
where, $P_L$ the laser power is in PW, $\rho$ in g/cc, $E_L$ is the laser energy in MJ and $R$ in cm.  
For achievable laser parameters: $P_L=0.5$ PW, $\rho=30$ mg/cc (foam material) and $E_L=0.5$ MJ, 
we have $P_{GW}\simeq6\times10^{-17}$ erg/s and $h_{GW}\simeq 10^{-39}$ for the detection distance of $R=$ 10~m (we believe
that a distance of a few meters is needed to protect the detector from strong broadband electromagnetic perturbations created in laser
interactions).

In the case of GW generation in the laser ablation zone, the rarefaction wave propagates with the sound velocity $C_s\simeq\sqrt{P_L/\rho_0}$. 
The expression for $Q_{zz}$ is similar to the shock wave problem, and $C_s$ have the same magnitude as $V_s$.
Hence, $P_{GW}$ and $h_{GW}$ have the values that are similar to the shock wave GW generation problem. 
In the laser matter interaction, the GW in the shock wave and in the rarefaction wave are produced at the same time 
and in the same space region. Then we can add these two contributions having, $P_{GW}\simeq 10^{-16}$ erg/s and $h_{GW}\simeq 4\times10^{-40}$.

Another GW source is a thin foil accelerated by a high ablation pressure produced 
by the laser heating and ablation\cite{AtzeniBook}. 
In this case, the quadrupolar mass momentum writes: $Q_{zz}(t)\simeq M z^2$, where $M$ is the initial mass foil.
We can assume that the mass ablated during the laser interaction is negligible, then the GW emitted power writes:
$P_{GW}\simeq G M^2 (\dot z \ddot z)^2/5c^5$ and $h_{GW}\simeq G M \dot z^2/c^4R$. With a high energy laser of a MJ  
class a velocity $\dot z \simeq 300$ km/s is achievable in $2$ ns for a foil mass about 2 mg. Then, $P_{GW} \simeq 5\times 10^{-19}$ erg/s and
$h_{GW}\simeq 1.5 \times10^{-40}$ with $R= 10$ m. 

HFGW could be produced with high power laser facilities dedicated to the inertial confinement 
fusion like the National Ignition Facility (NIF, USA), the Laser Megajoule (LMJ, France), or the European project for the inertial fusion energy HiPER \cite{Ribeyre09}. 
The laser driven implosion fusion can radiate HFGWs if the implosion of cryogenic deuterium-tritium (DT) 
micro-sphere would be asymmetric and produce a quadrupolar momentum $Q_{zz}$. In the case, $Q_{zz}\simeq\epsilon M z^2$, where $\epsilon
$ represents the non-symmetric part of the explosion. Such an asymmetric DT can be created by launching a bipolar shock\cite{Ribeyre09}.
 With this ignition scheme the asymmetry $\epsilon\simeq0.2$ is achievable.
The laser Megajoule can implode a mass around 0.2~mg, with a velocity $\dot z$ around $3\times10^8$ cm/s. 
The time scale of the explosion is around 20 ps. With these parameters, we have $P_{GW}\simeq 2\times10^{-14}$ erg/s and $h_{GW}\simeq3\times10^{-39}$
with $R=10$~m.  The fusion reactions produce in the central DT core high velocity jets, which radiate HFGW in 50 GHz 
domain during the plasma expansion of less than 100~ps.

Another possibility to generate GWs in the THz domain would be to use a high intensity picosecond laser pulse.
\cite{Tikhonchuk09}. A circular polarized pulse  with intensity $I_L\gtrsim 10^{21}$ W/cm$^2$ pushes a matter via the ponderomotive forceit 
it accelerates to the velocity $V_p=\sqrt{I_L/c \rho_0}$, which could be of the order of $10^9$ cm/s or more. 
The radiated power reads: $P_{GW}\simeq 5 \times 10^{-25} P_L^3 /\rho_0\Phi^2$ [erg/s], where $\Phi$ is the laser focal spot diameter (in cm) and $P_L$ in PW. Moreover, 
the metric perturbation writes: $h_{GW}\simeq 8\times10^{-40} P_L \tau_L/R$, where $P_L$ is in PW and $\tau_L$ in ps. A power about 7 PW during 1 ps, 
with a focal spot about 30 $\mu$m on a plastic target ($\rho_0=1$ g/cc), will be achievable on PetaWatt Class laser facilities (PETAL, NIF-ARC). 
With these parameters: $P_{GW}\simeq 2\times10^{-17}$ erg/s and $h_{GW}\simeq 6\times10^{-42}$ for a detection at the distance of 10 m.

\section{Conclusion}\label{aba:sec3}
Although all considered schemes have quite different geometries sizes and time scales, the generated GW powers and metric perturbations
 are not much different. This follows created by the observation that $h_{GW}$ from a point-like source can be estimated as $G {\cal E}/Rc^4$, where
${\cal E}$ is the available energy. With the maximum laser energy available today of 1 MJ, $h_{GW}$ cannot be larger at the distance of few
metters than $10^{-39}$. The noise detection level $(h_{GW})_{min} \simeq 10^{-30}$ given in Ref.~\citen{Rudenko03} is many order of magnitude higher. 
We conclude that available today laser sources are insufficient to generate HFGW on a detectable level. The limitation that we found for the point-like 
sources (with the size of the order of the emission wavelength) apply also to the sources of a larger size that might use interference effects to 
collimate emission in a certain solid angle. Although this may increase the intensity at the detector by one or two orders 
of magnitude, this do not affect the total emission power, which is still far away from the detection threshold.


\bibliographystyle{ws-procs975x65}
\bibliography{ws-pro-sample}

\end{document}